\PassOptionsToPackage{english}{babel}
\documentclass[aps, reprint,prl,superscriptaddress,nofootinbib]{revtex4-2}
\usepackage{dcolumn}
\usepackage[utf8]{inputenc}
\usepackage{amsmath}
\usepackage{amsfonts}
\usepackage{amssymb}
\usepackage{graphicx}
\usepackage{hyperref}
\usepackage{xcolor}
\usepackage{adjustbox}
\usepackage{comment}
\usepackage[toc,page]{appendix}
\usepackage[english,ngerman]{babel}
\newcommand{\ket}[1]{\ensuremath{|{#1}\rangle}}

\usepackage{tikz}
\usetikzlibrary{quantikz2}
\begin{document}
\selectlanguage{english}
\title{The Bose-Marletto-Vedral experiment with nanodiamond interferometers: an insight on entanglement detection}

\author{Giuseppe Di Pietra}
\email{giuseppe.dipietra@physics.ox.ac.uk}
\affiliation{Clarendon Laboratory, University of Oxford, Parks Road, Oxford OX1 3PU, United Kingdom}

\author{Fabrizio Piacentini}
\author{Ettore Bernardi}
\author{Ekaterina Moreva}
\author{Carmine Napoli}
\affiliation{INRIM, Strada delle Cacce 91, I-10135 Torino, Italy}

\author{Ivo Pietro Degiovanni}
\author{Marco Genovese}
\affiliation{INRIM, Strada delle Cacce 91, I-10135 Torino, Italy}
\affiliation{INFN, sez. di Torino, via P. Giuria 1, 10125 Torino, Italy}

\author{Vlatko Vedral}
\affiliation{Clarendon Laboratory, University of Oxford, Parks Road, Oxford OX1 3PU, United Kingdom}

\author{Chiara Marletto}
\affiliation{Clarendon Laboratory, University of Oxford, Parks Road, Oxford OX1 3PU, United Kingdom}

\date{\today}%

\allowdisplaybreaks
\begin{abstract}
Recently, it has been proposed a new method [\href{https://arxiv.org/abs/2405.21029}{arXiv:2405.21029}] to detect quantum gravity effects, based on generating gravitational entanglement between two nano-diamonds with Nitrogen-Vacancy defects, in a magnetically trapped configuration. Here we analyse in detail the proposed experimental setup, with a particular focus on implementing the detection of the gravitationally-induced entanglement using an optical readout based on measuring the position of the nano-diamonds and its complementary basis. We also summarise some of the key theoretical and experimental ideas on which this proposed scheme is based. 
\end{abstract}

\maketitle

\section{Introduction}
There is a long-standing debate in the scientific community about whether gravity should be quantised.
The fundamental incompatibilities between the two most successful contemporary physical theories -- general relativity and quantum theory -- have polarised the discussion.
Some believe that quantum theory cannot be applied beyond a certain scale \cite{karolyhazy_gravitation_1966,ghirardi_unified_1986,bassi_models_2013,hooft2022}, with gravity possibly being responsible for the collapse of macroscopic superposition, \cite{diosi_universal_1987,penrose_gravitys_1996}; others are either developing mathematical models to solve these incompatibilities, \cite{kibble_geometrization_1979,zwiebach_first_2004,rovelli_loop_2008,kiefer_quantum_2012,castellani_group_2022}, or providing theoretical arguments to show that, quoting Feynman, ``we are in trouble if we believe in quantum mechanics, but we do not quantise gravitational theory", \cite{rickles_role_2011,dewitt_global_2003,marletto_why_2017,vedral_are_2022}.

However, the physical and technological challenges in designing an experiment to \textit{directly} test quantum effects in gravity, as suggested by many proposals, pose a significant obstacle to confirming conclusively the quantum nature of gravity, \cite{boughn_aspects_2006,rothman_can_2006,dyson_is_2013,rov2024}.

Recently, a proposal has been put forward for a quantum gravity test, that promises to settle this debate by providing a witness of whether gravity is quantum, \cite{marletto_quantum-information_2024}. This test, also called the Bose-Marletto-Vedral (BMV) experiment, is based on witnessing the entangling power of gravity, \cite{bose_spin_2017,marletto_gravitationally_2017}. In particular, given two spatially separated masses $Q_A$ and $Q_B$ only \textit{locally} interacting with the gravitational field $M$, the idea is to detect gravitationally-induces entanglement (GIE) between $Q_A$ and $Q_B$: gravity can entangle the two masses only if it is non-classical. Interestingly, this test is based on an information-theoretic, indirect witness of non-classicality known as General Witness Theorem \cite{marletto_witnessing_2020} which provides a robust theoretical basis for GIE as in the BMV experiment to be a conclusive witness of gravity's quantum features.

\begin{figure}
    \centering
    \includegraphics[width=\linewidth]{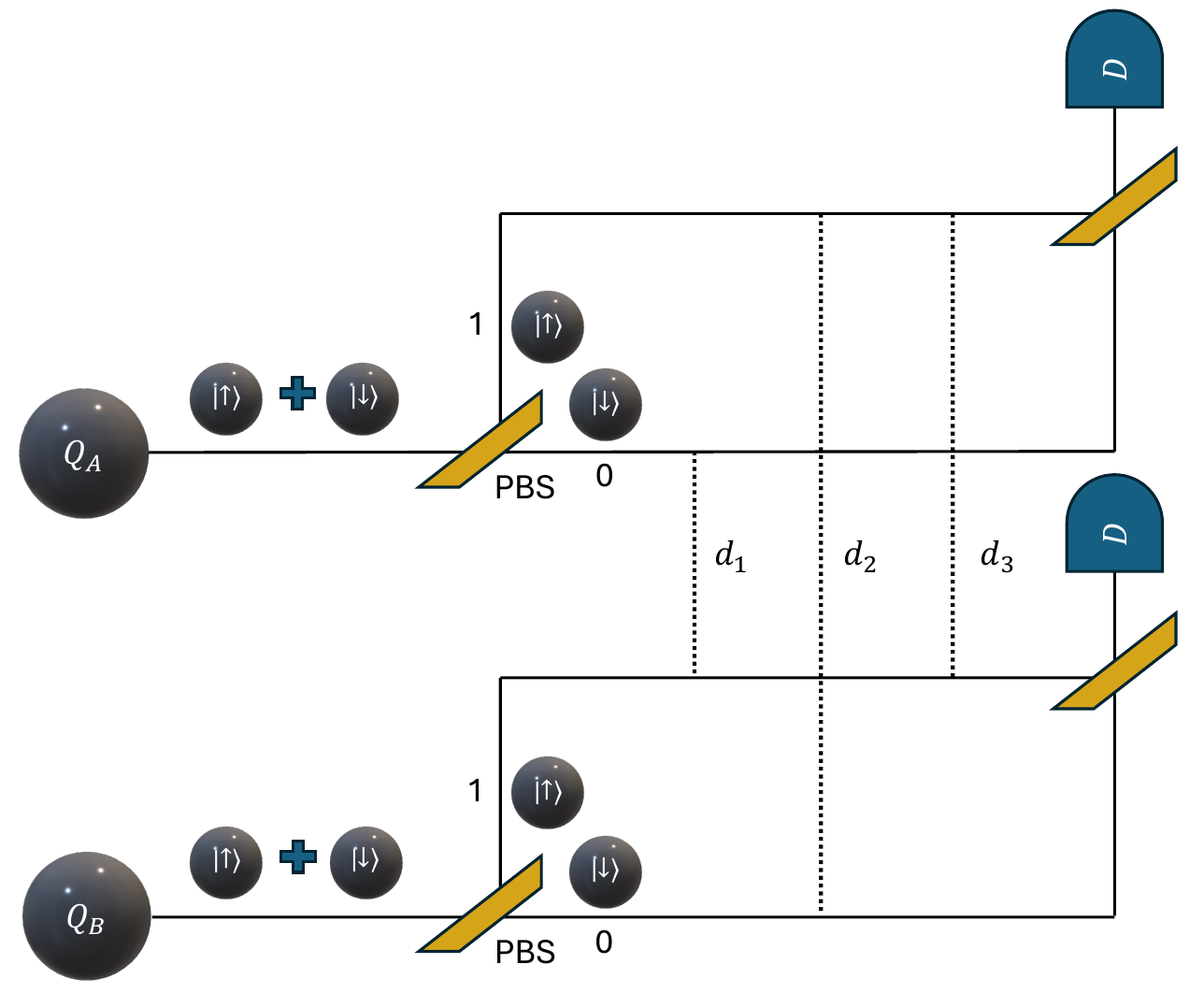}
    \caption{Schematic description of the BMV experiment. The two quantum probes $Q_A$ and $Q_B$, prepared in a superposition of their spins, undergo interference individually, interacting via gravity only. PBS labels the polarising beam splitter conditioned on the spin of the quantum probes, $0$ and $1$ the two paths involved in the interference, $D$ the detector following the second PBS, $d_1$, $d_2$ and $d_3$ are the distances between the upper arm of the first and the lower arm of the second interferometers, the lower arm of the first and the upper one of the second interferometers, and the upper arms of the interferometers respectively.}
    \label{fig:BMVscheme}
\end{figure}
Consider the simplified scenario in Fig. \ref{fig:BMVscheme}, \cite{marletto_gravitationally_2017}.
The two quantum systems $Q_A$ and $Q_B$ are mesoscopic systems of equal mass $m$ individually undergoing interference, with the interferometers located so that both masses are subject to the same Earth's gravitational field.
All direct interactions between $Q_A$ and $Q_B$ are excluded, and they couple only via gravity.
The first polarising beam splitter prepares each mass in the superposition $\ket{\psi}_i=\frac{1}{\sqrt{2}}\left(\ket{0}+\ket{1}\right)$, where $i=A,B$ and $0$ and $1$ label the two arms of each interferometer, depending on their spin: if the quantum probe's spin is $\ket{\uparrow}$, the PBS will reflect it along the path $1$, otherwise along the path $0$.
The gravitational interaction induces a distance-dependent phase $\phi_j\sim\frac{Gm^2t}{\hbar d_j}$ $(j=1,2,3)$ in each of the superposition branches. Hence at the end of the interferometry, the two masses are in the following state:
\begin{equation}
   \ket{\Psi}=\frac{1}{2}\left(\ket{0}\ket{0}+\ket{1}\ket{1}+e^{i\Delta\phi_1}\ket{0}\ket{1}+e^{i\Delta\phi_2}\ket{1}\ket{0}\right),
    \label{eq:finalstatePRL}
\end{equation}
characterized by the relative phases $\Delta\phi_1=\phi_1-\phi_3$ and $\Delta\phi_2=\phi_2-\phi_3$ associated with the closest and farthest massive superposition components.

Using linear quantum gravity -- which all the current quantum gravity proposals reduce to when Newtonian contributions are dominant, \cite{kiefer_quantum_2012} -- one can compute the phase differences $\Delta\phi_i$ generated by the gravitational interaction in $\ket{\Psi}$, \cite{bose_spin_2017, marletto_gravitationally_2017, marletto_when_2018}.
Consider a quantised gravitational field with a single polarisation described by the creation and annihilation operators $a^\dagger_k$ and $a_k$ respectively.
Here $k$ represents the wave number of the relevant mode of the gravitational field, which we shall first assume to be discrete for simplicity.
The general linearised Hamiltonian is $H_{int}^G=-\frac{1}{2}h_{\mu\nu}T^{\mu\nu}$, where $T^{\mu\nu}$ is the stress-energy tensor and $h_{\mu\nu}\propto\sum_{k}\frac{1}{\sqrt{\omega_k}}\left[a_ke^{ikx}+h.c.\right]$ is the perturbation of the metric tensor $g_{\mu\nu}$ away from the Minkowski flat spacetime.
In the BMV experiment, $Q_A$ and $Q_B$ are non-relativistic, so the stress-energy tensor simplifies to $T_{00}=m$.
The total Hamiltonian describing the two masses, the gravitational field and their interaction thus reads:
\begin{align}
    H=&mc^2\left(b^\dagger_Ab_A+b^\dagger_Bb_B\right)+\sum_k\hbar\omega_ka^\dagger_ka_k & \nonumber \\ & -\sum_{k,n\in\{A,B\}}\hbar g_kb^\dagger_nb_n\left(a_ke^{ikx_n}+a^\dagger_ke^{-ikx_n}\right)
    \label{eq:HamiltonianLG}
\end{align}
where $b_i$ and $b^\dagger_i$ $(i=A,B)$ are the annihilation and creation operators, respectively, describing the two equal masses $Q_A$ and $Q_B$.
The gravitation-matter coupling constant is $g_k=mc\sqrt{\frac{2\pi G}{\hbar \omega_k V}}$, where $V$ is the relevant volume of quantisation. One can solve exactly the dynamics generated by $H$ on this system, considering the field in the vacuum state and the two equal masses $m$ at positions $x_A$ and $x_B$ respectively: $e^{iHt}\ket{m}\ket{m}\ket{0}=exp\left\{\hbar \sum_k V(k) t\right\}\ket{m}\ket{m}\ket{\sum_k\frac{g}{\omega_k}\left(e^{-ikx_A}+e^{ikx_B}\right)}$, where $V(k)=\frac{g_k^2}{2\omega_k}\left[1+2\cos{\left(-i k(x_B-x_A)\right)}\right]$.
If the two masses are each in a spatial superposition, the Hamiltonian in Eq. \eqref{eq:HamiltonianLG} generates entanglement between them.
To compute the quasi-Newtonian phase that can generate the relevant entanglement, one can focus on the continuum version of the position-dependent part of $V(k)$:
\begin{equation}
    Re\left\{V\int dk \frac{4\pi Gm^2}{\hbar k^2 V}e^{-ik(x_B-x_A)}\right\}=\frac{Gm^2}{\hbar(x_B-x_A)}.
\end{equation}

A successful BMV experiment would rule out all classical theories of gravity that comply with the general assumptions on which its theoretical argument is based -- which we shall review in the next Section -- such as general relativity, and quantum field theory in curved spacetime, \cite{dewitt_quantum_1975,birrell_quantum_1982,biswas_towards_2012,tilloy_sourcing_2016}.
Moreover, it will exclude all the theories where gravity is either considered responsible for the collapse of the wave function at a certain scale, \cite{diosi_universal_1987,ghirardi_unified_1986,penrose_gravitys_1996}, or an entanglement breaking channel leading to decoherence, \cite{kafri_classical_2014}.

The groundbreaking implications of this novel witness of non-classicality for gravity make it crucial to devise a feasible experimental test with current technologies.
A recently proposed experimental scheme \cite{vicentini_table-top_2024} may represent a cornerstone in this challenge.
Two semi-trapped single nitrogen-vacancy (NV) centre nanodiamonds (NDs) are prepared in a superposition of their spin components $\sigma_x$.
The two superposition components of each ND are then spatially separated as in Fig. \ref{fig:BMVscheme} by a magnetic field gradient $B'$, playing the role of the initial PBS in the two interferometers.
Afterwards, the magnetic field gradient $B'$ is switched off, and their interaction is mediated solely by gravity: the mismatch between the phases accumulated on the different paths should then be responsible for their entanglement, as predicted by the non-classicality witness.
To measure the entanglement, in the final part of the protocol the $B'$ gradient is turned on again and the superposition components of each ND are recombined, realizing the scenario after the final PBS in the interferometers of Fig. \ref{fig:BMVscheme}, and subsequently a Ramsey-like readout of the NV centres spins is performed.

After reviewing this novel experimental scheme, we shall discuss which observables can be measured in this setup to witness the entanglement between the two NDs in the interferometers.
In fact, the entanglement is generated on the paths' degrees of freedom of the two NDs, rather than on the spin.
This makes the interferometers employed different from standard Stern-Gerlach interferometry, determining a crucial advantage of this scheme compared to others previously proposed in the literature.

The contribution coming from this analysis finalises the most advanced and promising experimental scheme to perform the BMV experiment while opening the way to potential other applications of the entanglement-based witness of non-classicality to other dynamics-agnostic systems, relevant for applications in e.g., quantum biology.

\section{The BMV experiment: a witness of non-classicality}
\label{sec:theory}
Consider a composite ``hybrid" system made of a quantum sector $Q$, e.g., a qubit, and an \textit{unknown} sector $M$, which may or may not obey quantum theory.
$M$ can be a complex system characterised by untractable dynamics, e.g., a biological system, or a system for which quantum theory as we currently understand it may break down, the most important example being gravity.
What information can one extract about $M$ in this hybrid system, \textit{without} the possibility of directly accessing it?
For example, can one infer whether $M$ has quantum properties?

For a conclusive investigation of the unknown system $M$, it is crucial to derive an argument that avoids any \textit{a priori} assumptions on it, and just relies on general information-theoretic principles, i.e., \cite{deutsch_constructor_2015}:
\begin{enumerate}
\item the Locality principle;
\item the principle of Interoperability of Information.
\end{enumerate}
These two principles are general in the sense that they are formulated independently of particular dynamics and they are obeyed by the best currently known descriptions of reality - quantum physics and general relativity; it is also reasonable to expect them to be obeyed by post-quantum theories too.

Equipped with a theoretical argument based on these principles, and exploiting the known sector $Q$, one could conclusively investigate a more general property of $M$: \textit{non-classicality}.
A physical system is \textit{non-classical} when it must be described by at least two non-compatible variables.
Informally, two variables are non-compatible when they cannot be measured to an arbitrarily high degree of accuracy by the same measuring device.
This definition generalises what in quantum theory one calls ``non-commuting" variables, with the advantage of not relying on any background dynamical assumption.\\
\begin{figure}
    \centering
    \includegraphics[width=\linewidth]{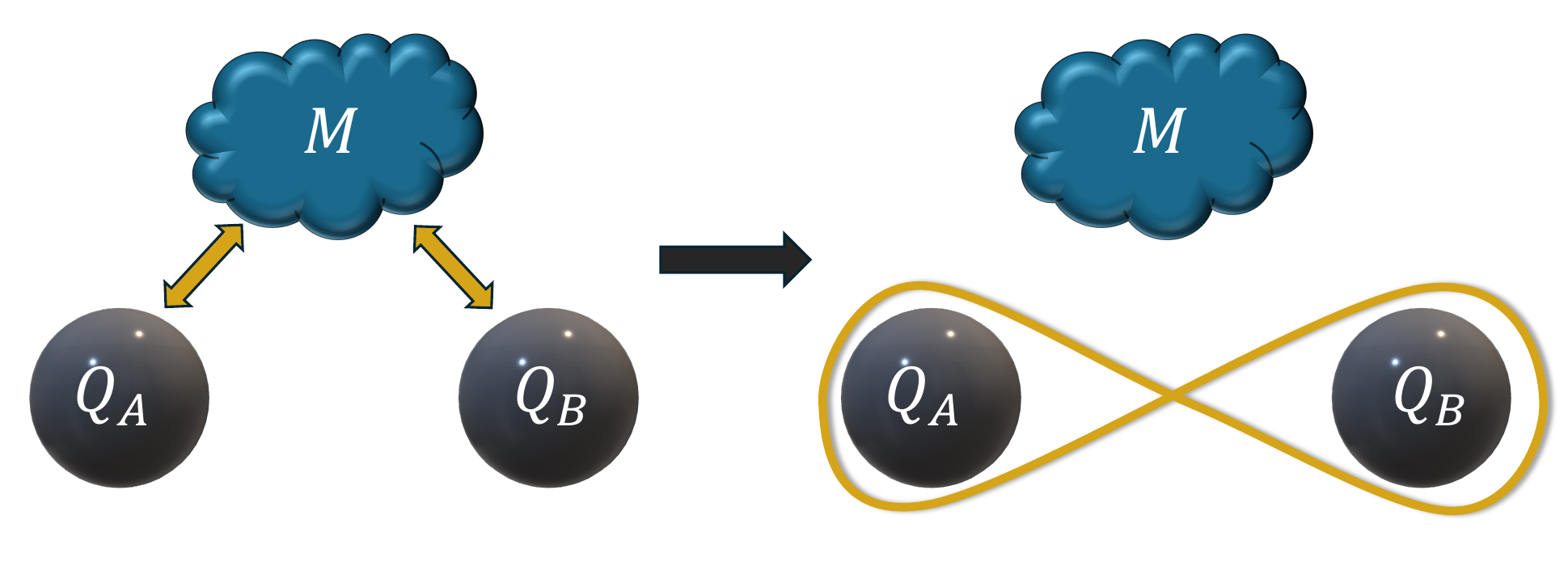}
    \caption{Visual representation of the entanglement-based witness of non-classicality. The two quantum probes $Q_A$ and $Q_B$ are spatially separated and can interact \textit{locally} with the system under investigation $M$. Under these assumptions, the creation of $M$-mediated entanglement between $Q_A$ and $Q_B$ would be a witness of $M$'s non-classicality.}
    \label{fig:BMVschemegeneral}
\end{figure}
The theoretical argument needed for this purpose has recently been proposed in \cite{marletto_witnessing_2020} and is known as the ``entanglement-based witness of non-classicality", or the General Witness Theorem: if the unknown system $M$ can mediate entanglement between two space-like separated quantum probes $Q_A$ and $Q_B$ by \textit{local interactions} only, then $M$ must be non-classical (see Fig. \ref{fig:BMVschemegeneral}).
``Local interactions" means here that $Q_A$ and $Q_B$ are separately coupled to $M$, but cannot interact directly.

The BMV experiment \cite{bose_spin_2017,marletto_gravitationally_2017} applies this witness within the formalism of quantum theory.
Despite the extra assumption of the quantum formalism, the argument stays general in the sense that it can be applied to any continuous or discrete system, independently of any particular model describing it and of the exact details of the dynamics, see \cite{marletto_quantum-information_2024,marletto_why_2017,marletto_witnessing_2017}.
However, when referring to ``non-classicality", one implies that the two non-compatible variables are in fact \textit{non-commuting} variables.

We shall now show that the BMV experiment is indeed a witness of the non-classicality of $M$ assuming, for simplicity, that $Q_A$ and $Q_B$ are two qubits.
Let $\hat{q}^{(A)}\coloneq \left(\sigma_x\otimes \mathrm{I}_{B,M},\sigma_y\otimes \mathrm{I}_{B,M},\sigma_z\otimes \mathrm{I}_{B,M}\right)$ be the vector of descriptors $q_k^{(A)}$ of the qubit $Q_A$ in the Heisenberg description of quantum theory \cite{deutsch_information_2000}.
Here $\sigma_k, k=x,y,z$, are the Pauli operators and $\mathrm{I}_{B,M}=\mathrm{I}_B\otimes \mathrm{I}_M$ is the identity operator on the qubit $Q_B$ and the unknown system $M$.
Similarly, $\hat{q}^{(B)}$ shall be the vector of descriptors of the qubit $Q_B$.

The proof goes by contradiction, thus we shall assume $M$ being a \textit{classical} system.
According to the definition of non-classicality stated above, this means it can be described by a single variable $T$.
Without loss of generality, we assume $T$ to be a binary variable and we represent it as an operator $q_z^{(M)}\coloneq \mathrm{I}_{A,B}\otimes \sigma_z$, where $\mathrm{I}_{A,B}$ is the identity operator on the two quantum probes.
The three subsystems are initially in a \textit{separable} state, meaning they are not entangled.
Following assumptions (1) and (2), $Q_A$ interacts with $M$ and, separately, $Q_B$ interacts with $M$ too; after these interactions, we assume that $Q_A$ and $Q_B$ get entangled.
However, the most general state of the global system one can get with a classical $M$ and local interactions is:
\begin{align}
    \rho=\frac{1}{4} & \left(\mathrm{I}+\Vec{r}_A\cdot\hat{q}^{(A)}+\Vec{r}_B\cdot\hat{q}^{(B)}+ s_zq_z^{(M)}+ \nonumber \right. & \\ & \left. + \Vec{t}_A\cdot\hat{q}^{(A)}q_z^{(M)} + \Vec{t}_B\cdot\hat{q}^{(B)}q_z^{(M)} \right),
    \label{eq:generalstate}
\end{align}
where: $\Vec{r}_A, \Vec{r}_B, \Vec{t}_A, \Vec{t}_B$ are real-valued vectors and $s_z \in \mathbb{R}$; $\mathrm{I}=\mathrm{I}_A\otimes\mathrm{I}_B\otimes\mathrm{I}_M$ is the identity on the Hilbert space of the global system $Q_A\oplus Q_B \oplus M$.
When interpreted as a two-qubit state, this state is separable, which contradicts the assumption of $Q_A$ and $Q_B$ being entangled.
The contradiction arises from the classicality of $M$, i.e., from the absence of a second variable for $M$ non-commuting with $T$, which concludes the proof: a classical mediator $M$ cannot entangle two quantum probes by local interactions only, if they are initially disentangled.

It is important to note that the assumption of Locality (1) is essential for the witness to hold true: a direct interaction between $Q_A$ and $Q_B$ could be responsible for the entanglement generation irrespectively of the non-classical nature of $M$, making the witness non-conclusive.
Indeed, if this were the case, Eq. \eqref{eq:generalstate} would acquire a term proportional to $\hat{q}^{(A)}\hat{q}^{(B)}$, making the two-qubit state non-separable.
Assumption (2) is instead guaranteeing that $M$ can mediate the interactions between $Q_A$ and $Q_B$.
Since we can copy the information from one system to the other, $M$ can transfer information from $Q_A$ to $Q_B$ and thus entangle them.
For this to be possible \textit{locally}, $M$ needs two non-commuting variables: first, the two probes become entangled with $M$ using one of its variables; then, this entanglement is induced to the two probes interacting with the second variable of $M$.
Thus, one can conclude that such a scenario is a direct consequence of the assumptions (1) and (2): the former implies that any local operation on the subsystem $Q_A\oplus M$ cannot affect $Q_B$ (and vice-versa), while the latter implies that $M$ can mediate interactions between $Q_A$ and $Q_B$.
Moreover, the witness is a \textit{sufficient} condition for the non-classicality of $M$, which implies that not observing the entanglement generation between the probes $Q_A$ and $Q_B$ given the assumptions (1) and (2), and eventually the formalism of quantum theory, makes the test inconclusive on $M$.

\section{Nanodiamond interferometers: a table-top setup for the BMV experiment}
The powerful theoretical implications of the argument underlying the witness, discussed in the previous Section, go hand in hand with a challenging experimental implementation realizing the scheme in Fig. \ref{fig:BMVscheme}.
The main challenges are:
\begin{enumerate}
  \item Excluding the direct interactions between the two quantum probes $Q_A$ and $Q_B$ such as Casimir-Polder and electromagnetic interactions; i.e., enforcing the locality assumption;
  \item Ensuring that the mediator of the entanglement is the gravitational field, not other external sources, e.g., electromagnetic fields.
\end{enumerate}

A recently proposed experimental scheme \cite{vicentini_table-top_2024}, based on a table-top interferometric setup for NDs, introduces a paradigm shift that promises a significant advance towards the definitive implementation of the BMV experiment.
In this Section, we shall review this proposal and compare it with existing ones in the literature, to show how this experiment promises to solve the aforementioned experimental challenges.

This proposal involves two single-NV-centre NDs of mass $m\sim10^{-14}$ kg, prepared in a superposition of their spin component $\sigma_x$'s eigenstates $\ket{\uparrow}$ and $\ket{\downarrow}$, i.e. $\ket{\psi}=\frac{1}{\sqrt{2}}\left(\ket{\uparrow}+\ket{\downarrow}\right)$.
They are constrained in the $y$ and $z$ directions by an anti-Helmholtz configured coil \cite{jansen_manipulating_2020}, but free to move along the $x$ direction. Doughnut-shaped optical tweezers control the dynamics on the $x-y$ plane, reducing the light absorption by the semi-trapped NDs, potentially causing severe thermal damage in a vacuum, \cite{geiselmann_three-dimensional_2013,zhou_optical_2017}.
The two NDs are separated by a distance $d\sim300$ $\mu$m along the $x$ direction, at a temperature $T\lesssim4$ K and with low pressure ($\lesssim10^{-16}$ bar).
These conditions are required to suppress decoherence and avoid direct interactions which could invalidate the non-classicality witness.
In particular, to ensure that gravity is the only mediator contributing to entanglement between the two probes, a limit must be imposed to the minimum distance $d_1$ between the closest NDs superposition components, in order to exclude static and dynamical Casimir-Polder interaction, \cite{bose_spin_2017,barcellona_dynamical_2016,van_de_kamp_quantum_2020}.
To ensure a Casimir-Polder interaction one order of magnitude below the gravitational one, it was deemed that, considering the NDs density and dielectric coefficient, such a limit should be $d_1\gtrsim200$ $\mu$m, \cite{bose_spin_2017,van_de_kamp_quantum_2020}.
Furthermore, the neutrality of the quantum probes prevents electromagnetic interaction, \cite{marletto_gravitationally_2017}.
This is relevant to solving the experimental challenge $(1)$ listed above.

To spatially separate the two superposition components in $\ket{\psi}$, a magnetic field gradient $B'$ is applied along the $x$ direction.
This determines oscillatory dynamics for the two $\sigma_x$ components, with the equilibrium position dependent on the bias (static) magnetic field $B_0$.
The subsequent asymmetry between the two trajectories of each ND manifests in a phase difference accumulation $\Delta\phi$ \cite{vicentini_table-top_2024}, which can be characterised using Ramsey-like interferometry \cite{vandersypen_nmr_2005} when the two superposition components recombine in the original ND position.
The magnetic field gradient is switched on for the time needed for the two branches of each ND delocalized superposition to reach maximal separation; there, the two superposition components of each ND present null speed, hence turning off $B'$ would result in the superposition components keeping their position, minimizing the possibility of losing them because of drifts in the $x$ direction.

The spatial superpositions will take the key role in the interference experiment, meaning that the entanglement is generated on the position degree of freedom (while the spin is a label that is used to create the spatial superposition and then undo it when the interferometer is closed).
It is, therefore, crucial to ensure that the quantum coherence of each spatial superposition is maintained long enough for the gravitational field to entangle the different paths, which is of the order of a second, \cite{bose_spin_2017}.
This can be accomplished by introducing a dynamical decoupling (DD) mechanism \cite{pedernales_motional_2020}.
DD consists of a train of $N$ microwave $\pi$ pulses, which flip the sign of the spin component $\sigma_x$ in each branch at a frequency $\omega_{DD}=N\omega$, where $\omega=B'\sqrt{\frac{\chi}{\mu_0 \rho_{ND}}}$ is the oscillation frequency dictated by the magnetic field gradient $B'$ acting on a ND with density $\rho_{ND}$, magnetic susceptibility $\chi$, in the vacuum ($\mu_0$ being the vacuum magnetic permeability).
It should be stressed that the DD implementation in this setup requires a single microwave antenna, while a free-falling/travelling setup would need more than $10^4$ microwave antennas synchronised with the ND travelling speed, as in the proposal of Ref. \cite{bose_spin_2017}.
A simultaneous flipping of the magnetic field gradient $B'$ is required to maintain the dynamics time-invariant and make the two superposition components' trajectories symmetrical to the origin, as when $B_0=0$.
Thus, the DD mechanism in this setup makes the spatial superposition of the NDs \textit{insensitive} to residual static force fields which may be present in the experiment, with a remarkable step forward in solving the experimental challenge (2).
Moreover, to flip $B'$ in the DD mechanism one can employ the same anti-Helmholtz coil used for the semi-confinement of the two NDs.
This determines an optimal and versatile usage of the technological resources needed for the experiment in the semi-trapped configuration, reducing the experimental complexity while enhancing its feasibility and precision. This represents one of the key advantages of this proposal.

At this point, the magnetic field gradient $B'$ is switched off, so that only gravitational interactions between the different branches of the two superpositions remain.
During this step, whose duration must be chosen adequately for the desiderate effect to be visible, the gravitational field entangles the two NDs along the different paths.
To conclude the interference experiment, $B'$ is switched on again and the superposition of each ND is recombined.
Then, one performs a single-shot measurement of the global spin state of the two NDs to detect GIE (see the following Section). If GIE is detected, then the BMV experiment is successful in witnessing the non-classicality of the gravitational field.

It is important to note that, contrary to previous proposals, in this scheme, the two NDs are not lost after the measurement; indeed, they can be re-initialised in their initial state for another experimental run, without having to find, characterise and trap new samples, significantly reducing noise due to unavoidable fabrication mismatch among the different ND samples exploited in other schemes \cite{bose_spin_2017,wood_spin_2022} (not to mention the dramatic reduction of the time needed for the experiment completion).
Furthermore, a semi-confined configuration for the ND interferometers determines a more compact and controllable experimental setup with respect to the previous proposals based on free-falling/travelling schemes \cite{wan_free_2016,wood_spin_2022,marletto_gravitationally_2017,bose_spin_2017}.
The latter, in fact, require a low temperature, high vacuum magnetic structure $\gtrsim10$ meters long, with microfabricated structures in the $\mu m$ scale, that are challenging and expensive to construct, maintain and control; the DD mechanism would also be much more challenging to implement, as it would require several ($\sim10^4$) microwave antennas distributed along the whole structure of the experiment.
Finally, the fact that, in free-falling schemes, the NDs are lost at the end of each experimental run makes the repetition of the experiment extremely time-consuming (since the ND fabrication, characterisation and trapping tasks are the most time-consuming ones), more expensive and highly subject to external noise, both due to the fabrication of the samples and the fragility of the experimental setup.

\section{How to detect gravitationally induced entanglement in this setup?}
Once the spatial superposition of each ND is recombined, one should measure the global spin state of the quantum probes to certify the eventual entanglement induced by gravity.
To this end, it is important first to distinguish between the role of the spin and the paths' degrees of freedom. 
In fact, the entanglement witness must be performed on the path degrees of freedom, with the spin becoming simply a label for the different paths. 
Therefore, in the following, we shall separate the spin degree of freedom $\ket{\uparrow}$ or $\ket{\downarrow}$, and the path degree of freedom $\ket{0}$ or $\ket{1}$ in the relative state of each mass $Q_A$ and $Q_B$.

The entanglement witness we are interested in is of the kind $X_AZ_B+Z_AX_B$, where $Z_i$ and $X_i$ ($i=A,B$) indicate, respectively, the $Z$ and $X$ Pauli bases in the path degree of freedom \cite{marletto_when_2018}: $Z$ represents the discretised path of each ND; $X$ is its conjugate observable. 
To accumulate sufficient data and experimentally measure the expectation value of $X_AZ_B+Z_AX_B$, we can proceed in at least two ways to satisfy different experimental needs, as follows.
Let us assume here the best-case scenario where the phase gives us maximal entanglement between the NV centres.
Consider the maximally entangled state of the two NDs:
\begin{align}
\ket{\Phi}=\frac{1}{2}&\left [\ket{\uparrow}_A\ket{0}_A(\ket{\uparrow}\ket{0}+\ket{\downarrow}\ket{1})_B \right. \nonumber & \\ & \left. +\ket{\downarrow}_A\ket{1}_A(\ket{\uparrow}\ket{0}-\ket{\downarrow}\ket{1})_B\right].
\end{align}

The first strategy to measure $Z_AX_B$ can be summarised as follows:
\begin{itemize}
    \item First, measure the path of the first ND, $Q_A$. In this context, this corresponds to measuring $Z_A$. This can be achieved by measuring the NV-centre spin by means of the single-shot optically-detected magnetic resonance (ODMR) technique \cite{wrach97,Suter2017}, which serves as a proxy for measuring position. So if one observes $\sigma_x^{(A)}=\;\uparrow$, $Q_B$'s relative state is 
    $$\ket{\Phi}^\uparrow_B=\frac{1}{\sqrt2}(\ket{\uparrow}\ket{0}+\ket{\downarrow}\ket{1}).$$ If one observes $\sigma_x^{(A)}=\;\downarrow$, $Q_B$'s relative state is $$\ket{\Phi}^\downarrow_B=\frac{1}{\sqrt2}(\ket{\uparrow}\ket{0}-\ket{\downarrow}\ket{1}).$$
    \item Then, to measure the complementary basis $X_B$ to the position of the second ND, $Q_B$, one can apply the same magnetic field gradient $B'$ that was employed to create the spatial superpositions (corresponding to the PBS element of Fig. \ref{fig:BMVscheme}); this recombines the paths of the ND, thus disentangling the path and the spin degrees of freedom. At the end of this operation, one switches off the gradient $B'$ and then measures the spin value of the NV centre in the $x$ direction. The combined action of these two operations is equivalent to measuring $X_B$ on the ND. Hence, this measurement will either give a $+1$ result, corresponding to the NV centre being in the state $\frac{1}{\sqrt2}(\ket{\uparrow}+\ket{\downarrow})$, or $-1$,   corresponding to the NV centre being in the state $\frac{1}{\sqrt2}(\ket{\uparrow}-\ket{\downarrow})$. 
\end{itemize}
The same analysis applies, by symmetry swapping the roles of $Q_A$ and $Q_B$, to measure $X_AZ_B$.

The second strategy to measure $Z_AX_B$ instead goes as follows. One first applies the magnetic field gradient $B'$ to recombine the paths of the two NDs. Since this acts as a PBS (see Fig.\ref{fig:BMVscheme}), $Q_A$ and $Q_B$ will both exit the interferometers along the path $\ket{0}$, so that the state at the entrance of the detector $D$ will be:
\begin{align} 
    \ket{\Phi'}=\frac{1}{2}&\Big[\ket{\uparrow}_A\left(\ket{\uparrow}+\ket{\downarrow}\right)_B \nonumber & \\ &  + \ket{\downarrow}_A\left(\ket{\uparrow}-\ket{\downarrow}\right)_B \Big] \ket{0}_A\ket{0}_B
    \label{eq:secondstrategy}
\end{align}
Eq. \eqref{eq:secondstrategy} shows a factorisation of the path degree of freedom of both the NDs, but still entanglement in the spin degree of freedom. Thus, in this case, $Z$ would be the observable with eigenstates $\{\ket{\uparrow},\ket{\downarrow}\}$ and $X$ the conjugate observable with eigenstates $\{\frac{1}{\sqrt{2}}(\ket{\uparrow}+\ket{\downarrow}),\frac{1}{\sqrt{2}}(\ket{\uparrow}-\ket{\downarrow})\}$.
The following steps can be then implemented:
\begin{itemize}
    \item First, using the ODMR technique mentioned above, one measures $Q_A$'s spin degrees of freedom to get what in our notation we call $Z_A$. If one observes $\sigma_x^{(A)}=\;\uparrow$, then $Q_B$'s state is $$\ket{\Phi'}^{\uparrow}_B=\frac{1}{\sqrt{2}}(\ket{\uparrow}+\ket{\downarrow})\ket{0}.$$ If $\sigma_x^{(A)}=\; \downarrow$ is observed, then $Q_B$'s relative state would be $$\ket{\Phi'}^{\downarrow}_B=\frac{1}{\sqrt{2}}(\ket{\uparrow}-\ket{\downarrow})\ket{0};$$
    \item Then, as described in the previous strategy, one measures $Q_B$'s spin along the $x$ direction, to measure $X_B$. In the first case, one would get $+1$ as a result of this measure, in the second, one would get $-1$.
\end{itemize}
Once again, a symmetrical analysis applies by swapping the roles of $Q_A$ and $Q_B$.

These two theoretical strategies show the interesting interplay between the spin and the position degrees of freedom to measure the entanglement witness in this setup. Moreover, they highlight the versatility of this scheme, which allows for multiple measurement procedures of the same entanglement witness, to comply with different experimental needs and help enhance its feasibility.

On a side note, the possibility of measuring the single phases acquired by $Q_A$ and $Q_B$'s superposition components at the end of the interference process could reveal the gravitational field power of inducing the quantum coherent evolution of the two probes' spin degrees of freedom, on top of its entangling power. 
Despite being less strong on the theoretical ground, thus not enough to conclude on the quantum nature of gravity alone, one could exploit a recently proposed ``temporal" witness of non-classicality, \cite{di_pietra_temporal_2023}, to confirm the results obtained from the BMV experiment as presented in this work. The argument goes as follows: if the system under investigation $M$ can induce the quantum coherent evolution of a quantum probe $Q$ conserving a global quantity on the system $Q\oplus M$, then it must be non-classical in the sense discussed in previous Sections. Thus, with a single experimental setup, one can also have an independent way of confirming the quantum nature of gravity.

\section{Conclusion}
The BMV experiment \cite{bose_spin_2017,marletto_gravitationally_2017} is one of the most promising approaches to conclusively testing the quantum nature of gravity. The strength of its theoretical foundations is accompanied by a challenging experimental implementation. Our experimental scheme, based on ND interferometers, is one of the most promising ways of implementing this proposal. In this paper, we have discussed in detail how to detect GIE with this novel setup, stressing that the quantum correlations are generated on the path degrees of freedom rather than on the spin, making this setup subtly different from a standard Stern-Gerlach interferometer. The advantages of this scheme compared to previously proposed ones based on free-falling mechanisms can be summarised in four main points: 1) simpler experimental setup, based on magnetic semi-trapping, which can be achieved on a table-top scale without the need for large-scale infrastructure; 2) improved control over the quantum probes, thanks to the compact anti-Helmholtz coil and doughnut-shaped optical tweezers; 3) enhanced robustness against external noise and perturbations, such as residual static fields; 4) more efficient use of resources, as the employed components can be used for different tasks and the sampled NDs can be used for multiple experimental runs.

We have discussed how to detect the entanglement in these novel interferometers at the theoretical level, stressing that the quantum correlations are generated on the paths' degrees of freedom rather than on the spin ones, and suggesting multiple strategies to measure entanglement witnesses of the kind $Z_AX_B+X_AZ_B$. This allows us to dispose of a wider set of tools to employ to optimise the efficiency of the experiment we aim to implement when it comes to measuring the entanglement generated by the gravitational field on the two space-like separated NDs $Q_A$ and $Q_B$.

The BMV experiment has never been this close to being realized.

\section{Acknowledgments}
G.D.P. thanks the Clarendon Fund and the Oxford-Thatcher Graduate Scholarship for supporting this research.
This work was funded by the Intesa San Paolo Foundation under the Trapezio project "QuteNoise". 
This research was made possible through the generous support of the Gordon and Betty Moore Foundation, the Eutopia Foundation and of the ID 62312 grant from the John Templeton Foundation, as part of the \href{https://www.templeton.org/grant/the-quantuminformation-structure-ofspacetime-qiss-second-phase}{‘The Quantum Information Structure of Spacetime’ Project (QISS)}. The opinions expressed in this project/publication are those of the author(s) and do not necessarily reflect the views of the John Templeton Foundation.

\bibliographystyle{apsrev4-2}
\bibliography{bibliography}

\end{document}